\documentclass[sigconf]{acmart}

\renewcommand\footnotetextcopyrightpermission[1]{} 
\makeatletter
\renewcommand\@formatdoi[1]{\ignorespaces}
\makeatother

\usepackage{booktabs} 
\usepackage{epstopdf} 

\usepackage{subfigure}
\usepackage{multirow}  
\usepackage{epigraph}
\usepackage{booktabs}
\usepackage{float}
\usepackage{morefloats}
\usepackage{url}
\usepackage{graphicx}

\usepackage[utf8]{inputenc}

\setcopyright{none}

\acmDOI{}

\acmISBN{}

\acmConference{preprint}
\acmYear{2018}
\copyrightyear{}

\acmPrice{}

\begin{document}
\title{Collaborative Filtering vs. Content-Based Filtering:\\differences and similarities}

\author{Rafael Glauber and Angelo Loula}
\affiliation{%
  \institution{Intelligent and Cognitive Systems Lab (LASIC)}
  \streetaddress{State University of Feira de Santana (UEFS)}
  \city{Feira de Santana, Bahia, Brasil} 
}
\email{{rafaelglauber, angelocl}@ecomp.uefs.br}

\renewcommand{\shortauthors}{Glauber and Loula}

\begin{abstract}
Recommendation Systems (SR) suggest items exploring user preferences, helping them with the information overload problem. Two approaches to SR have received more prominence, Collaborative Filtering, and Content-Based Filtering. Moreover, even though studies are indicating their advantages and disadvantages, few results empirically prove their characteristics, similarities, and differences. In this work, an experimental methodology is proposed to perform comparisons between recommendation algorithms for different approaches going beyond the ``precision of the predictions''. For the experiments, three algorithms of recommendation were tested: a baseline for Collaborative Filtration and two algorithms for Content-based Filtering that were developed for this evaluation. The experiments demonstrate the behavior of these systems in different data sets, its main characteristics and especially the complementary aspect of the two main approaches.
\end{abstract}

%
%



\maketitle

\section{Introduction}
\label{sec:introduction}

The large volume of information available on the Internet has made it difficult for users to retrieve content of interest. This problem called Information Overload has been the main object of research on Recommender Systems (RS) \cite{silva2017recstore, santana2017evaluating, silva2017evaluating, fernandes2017using}. Content-based filtering (CB) and collaborative filtering (CF) are the main approaches for building such system. However, several authors \cite{das2007, shardanand1995, goldberg1992, konstan2012} indicate limitations in both approaches. Among the most cited for the content-based approach are do not surprising the user and  not filtering based on subjective issues such as quality and style. And the need for a wide range of positive and negative evaluations to generate good recommendations is a limitation for collaborative approaches. However, these works lack comparative results that support the idea that an approach is more adequate or not to solve the problem of information overload considering a specific context.

This paper proposes a comparative evaluation between RS and describes a methodology that can support the identification of which approach is more or less adequate for a given problem. For this, new dimensions are explored on the recommendations of the two main approaches to contribute to this discussion. In addition, we propose two memory-based algorithms for content-based recommendation and we compare the similarity between the recommendations made between different algorithms.

The remainder of this study is organized as follows: in the Section~\ref{sec:relatec} have presented some studies related to the object of study of our work. In the Section~\ref{sec:proposal}, we present our proposal and the important details for the development of our work. Then, in the Section~\ref{sec:results}, we present our results. In the end, we present our conclusions and propose some developments of this work in the Section~\ref{sec:remarks}.

\section{Related Work}
\label{sec:relatec}

A project to build a RS should contain a step of system evaluation. This step aims to evaluate the behavior of the system while executing its task, to compare it with other solutions (if any) or to define the best configuration of its parameters before entering the operation phase. There are two categories of evaluation: system evaluation (offline) and user evaluation (online) \cite{shani2011, voorhees2002}. Our study is focused on the offline evaluation category. The offline evaluation method focuses on turning the recommendation task into an item classification prediction task using user rating history. For example, retrieving the list of items evaluated by a user removes a portion of these items so that the RS guesses the assessments assigned by the user. It is expected that any RS proposal should be submitted for evaluation before deployment for end users, minimizing as much as possible bad experiences with the recommender by system users. In RS studies, there are variations on the methodology applied for comparing SRs. For example, \cite{Bogers2008} presents a trial configuration that tests three different CFs that are evaluated with CiteULike usage data using the time sequence of the system usage data. \cite {bobadilla2011improving, di2012linked, beel2013comparative, wu2015fuzzy} are examples of works that evaluated and compared their algorithms in different contexts (different datasets) or even against other algorithms using measures like Precision, Recall and others.

The work of \cite {bellogin2011precision} discusses four methodologies for offline evaluation of RS and possible distortions that can occur while performing variations in the experimental methodology. This paper reinforces the idea that the evaluation stage is of great relevance in the design of a RS. In \cite {Said2014Comparative} given different performances of each of the three recommendation algorithms tested, in different dimensions (e.g. nDCG, RMSE, User coverage and Catalog coverage), researchers suggest that there is no golden rule to evaluate RS. This argument should not be viewed as a limitation and as long as it is fair, new proposals should be developed for different contexts as the authors suggest.

\section{Proposal}
\label{sec:proposal}

Our study focus on the system evaluation phase of a RS that performs the task of Finding Good Items \cite{herlocker2004}. We propose an evaluation protocol based on  principles adapted from other studies. We organize our evaluation protocol as follows:

\begin{itemize}
    \item Dataset -- We build datasets from different sources. We are seeking greater generalization of our method.
    \item Algorithm -- We propose two CB RS to compare with a typical CF RS.
    \item Evaluation protocol -- We define a method for offline assessment. At this point, we propose changes in coverage measures. Besides, we analyzed the recommendation lists measuring the similarity between results from algorithms.
\end{itemize}

\subsection{Datasets}

We use different recommendation contexts  through two datasets: movies and scientific articles. The MovieLens \footnote{http://www.movielens.umn.edu/} project provides user review datasets for RS searches. In this work, we used the \textit{MovieLens 1m Data Set} with 1 million ratings of 6,000 users in 4,000 movies and each user has at least 20 ratings. Originally, the dataset has little content for items (movies), only title, genre and year. To expand the content, we've built a web crawler to capture more data from IMDb \footnote {https://www.imdb.com/}, a collaborative online service for movies. CiteULike\footnote{http://www.citeulike.org/} is a web service to organize a library of scientific references for users, and user activity data on the system has been made available for research. Unlike the MovieLens, the dataset offers raw data with no data selection filter, so we selected users and activities from 2013. In CiteULike, the user is not invited to evaluate  items in the system. It is possible, however, to define user interest implicitly: any scientific article present in the user's library should be of interest to you. The two datasets used (MovieLens and CiteULike) are presented in Table~\ref{tab:dataset_com_filtro} in their final configuration.

  \begin{table}[ht!]
\centering
\small
\begin{tabular}{ l  c  c } 
\hline
\textbf{Property}          $\,$ & $\,$ \textbf{MovieLens -- 1m} $\,$ & $\,$ \textbf{CiteULike}\\ 
\hline
\#users        		      $\,$ & $\,$ 6038 	$\,$ & $\,$ 3000\\
\#items  			          $\,$ & $\,$ 3533 	$\,$ & $\,$ 32563\\
\#activities		          $\,$ & $\,$ 575279 	$\,$ & $\,$ 34037\\
\hline
\#items / \#users		$\,$ & $\,$ 0,5847 	$\,$ & $\,$ 10,8543\\
avg \#items by user	      $\,$ & $\,$ 95,2764 	$\,$ & $\,$ 11,3456\\
avg \#users by item	      $\,$ & $\,$ 162,8301 	$\,$ & $\,$ 1,0452\\
\hline
max \#items by user       $\,$ & $\,$ 1435	    $\,$ & $\,$ 947\\
max \#users by item       $\,$ & $\,$ 2853		$\,$ & $\,$ 20\\
\hline
min \#items by user       $\,$ & $\,$ 1	        $\,$ & $\,$ 2\\
min \#users by item       $\,$ & $\,$ 1 		$\,$ & $\,$ 1\\
\hline
esparsity $user \times item$ $\,$ & $\,$ 0,9730 	$\,$ & $\,$ 0,9996\\
\hline
\end{tabular}	
\caption{Statistics on the datasets (MovieLens and CiteULike) used in this study.}
\label{tab:dataset_com_filtro}
\end{table}

Each recovered dataset has distinct attributes, since the attributes for movies differ from the scientific publications. In the first set, with specific information about movies and series, the recovered data was Title, Synopsis, Keywords, Genre, Stars, Writers, Directors, User review, Release. In the second set, the attributes of the publications were Title, Abstract, Authors, Publication, Publisher, Source, Tags, Year.

\subsection{Recommender Algorithms}

All recommenders used in our experiment are memory based. Our social recommender is a baseline system described by the authors in \cite{su2009}. In the content-based recommendation, we developed two algorithms that use the Bag-of-Words representation and in the preprocessing step we perform the following filters: removal of stopwords and removal of terms with $DF=1$. The representation of the documents in the array ``attribute x value'' was done using weight $TF-IDF$ without normalization. The similarity calculation was through Cosine Similarity \cite{chowdhury2010introduction}. Our two algorithms are described as follows:

\textit{User Profile Aggregation} (UPA) is based on a term content filter. The model that represents the user's preferences is a set of words obtained from all user profile items. The rationale is that user interest can be modeled through the most relevant terms present in the items from the user profile. This algorithm can be subdivided into four main steps performed sequentially: create the representation model of the items through their contents; extract the 100 most relevant terms from the items present in the user profile to define his preferences; calculate the similarity between the relevant words and the other items in the system; retrieve the most similar items to compose the recommendation list.

\textit{Similarity of User Profile} (SUP) proposes that each item in the user profile is a voter, and the other items in the system are candidates to enter a recommendation list. A recommendation is made by determining the similarity between voters (user profile items) and candidates (other items of the system) and the similarity value between  voter and  candidate is used as a weight for the vote. The 50 most similar items to each item in the user profile are candidates and those with the most votes are retrieved and form a recommendation list for the user.

\subsection{Evaluation}

We evaluated the recommendations in configurations in which each user had 10 profiles removed (\textit{Given~10}). The remaining data was used as a training set. We defined that users must have at least 10 training items. A cross-validation procedure with \textit{10-fold} was performed to obtain estimates of the expected behavior of RS. The total data set is divided into 10 subsets and, each time, 9 subsets are used for training and 1 for testing.

\subsubsection{Measures}

\textit{Mean Average Precision} was chosen as our main evaluation measure for the referrals. This measure evaluates the recommendation list based on the position that the item considered relevant (in this case, removed from the user profile) is located. Hence the best results are obtained early in the list. In our evaluation, we established fixed sizes for the recommendation lists (10, 20, 30, 50 and 100) and thus the MAP score is limited to this configuration. Because of this we will call this measure throughout the work as \textit{MAP at K} (MAP@K).

In order to evaluate the ability of a recommender that performs the ``recommend good itens'' task in generating recommendations to every user, we use \textit{UCOV at K} or simply UCOV when convenient. We define this measure as:

\begin{equation}
User\ Coverage=\frac{1}{|U|}\times \sum_{u=1}^{|U|} \frac{|L_u|}{K}
\end{equation}

where $|L_u|$ is the size of the recommendation list generated for user $u$ e $U$ is the set of active users that should receive recommendation lists of size $K$.  This measure aims to analyze the ability of the tested algorithms in  accomplishing their task of composing a full list of recommendation for each user.

We also use \textit{Catalog Coverage} to evaluate the ability of the system to cover the entire catalog of items, taking into account recommendations for all users. In our experimental methodology, the recommenders perform the task of recovering good items for each active user only once and then the system is evaluated. In this way, based on the measure presented in \cite{ge2010}, CCOV at K, or just CCOV, for lists of maximum size $K$ was adapted in our experiment as:

\begin{equation}
Catalog\ Coverage=\frac{|L_{u=1}\bigcup L_{u=2}\bigcup...\bigcup L_{u=|U|}|}{|I|}
\end{equation}

where $I$ is the set of items available in the system, that is, the catalog, $L$ is the recommendation list that is displayed for each user $u \in U$, where $U$ is the set of active users.

In this study we propose to evaluate the similarity between the recommendation lists of the evaluated algorithms. In a prediction task, two algorithms may have the same prediction, but this result may mask a question: are the generated recommendation lists similar? Are all the predictions made by each of the approaches algorithms similar or different? By means of the recommendation task evaluated in this study, we propose to calculate the similarity among the recommendation lists using the Jaccard similarity. This procedure is performed by taking the recommendation list generated by each of the algorithms, two at a time, for the same active user. Finding similarity between recommendation lists is the starting point for our similarity analysis among recommenders. Then the proportions of correct hits generated by the RS are verified. In this analysis, we took two recommenders and separated the prediction hits into three distinct sets: the unique hits of algorithm A, the unique hits of algorithm B, and the intersection of the hits of A and B. The first set indicates which predictions are uniquely matched by the algorithm A. The second set is analogous to the first and the third is the list of accurate predictions by both algorithms. The analysis of these three sets indicates whether there is overlap of prediction hits in the recommendations generated by the algorithms. That is, if the correct predictions from an algorithm are included in the prediction lists of another algorithm, or if the algorithms have different prediction hits.

\section{Results}
\label{sec:results}

Our results are organized for each dataset used and the evaluation measures described in this study. First, we present the precision results and then the coverage results. In the end, we evaluated similarity and intersection between the algorithms.

\subsection{Precision}

When the ``recommending good items'' task is performed, hitting hidden items from the user profile in the first few places demonstrates an RS's ability to learn user preferences.

\subsubsection{Precision for MovieLens}

Table~\ref{tab:given10_map_movielens} contains results for the two recommendation algorithms based on textual content: User Profile Aggregation (UPA) and Similarity of User Profile (SUP). Results refer to the dataset MovieLens and the task used is to predict 10 items removed from the user profile (Given~10). The algorithm that presents the best results is the SUP when used \textit{All} the concatenated attributes. Besides, the concatenation of all attributes followed by the concatenation \textit{1+2+3} are the best attributes for precision of predictions, i.e. the more context is available the more process are the content based recommenders. Another aspect is the increase in MAP values when the lists grow. Nevertheless, this growth can not be considered relevant given the sizes of the lists 10 and 20, 20 and 30, 30 and 50 and 50 and 100. This suggests that the SUP algorithm, following a content-based approach, concentrates the correct hits in the beginning of the list.

\begin{table}[ht]
\centering
\small
\scalebox{0.65}{
\begin{tabular}{l c|c|| c|c|| c|c|| c|c|| c|c}
\toprule
& \multicolumn{10}{c}{\textbf{Given 10}} \\
\cmidrule(r){2-11}
& \multicolumn{2}{c|}{MAP@10} & \multicolumn{2}{|c|}{MAP@20} & \multicolumn{2}{|c|}{MAP@30} & \multicolumn{2}{|c}{MAP@50} & \multicolumn{2}{|c}{MAP@100}\\
\cmidrule(r){2-11}
& SUP & UPA & SUP & UPA & SUP & UPA & SUP & UPA & SUP & UPA \\
\cmidrule(r){2-11}
Title          & 0,0023 & 0,0003 & 0,0027 & 0,0006 & 0,0029 & 0,0009 & 0,0031 & 0,0013 & 0,0038 & 0,0018 \\
Summary         & 0,0107 & 0,0009 & 0,0114 & 0,0015 & 0,0118 & 0,0019 & 0,0123 & 0,0024 & 0,0129 & 0,0031 \\
Keyword        & 0,0059 & 0,0002 & 0,0064 & 0,0004 & 0,0067 & 0,0007 & 0,0071 & 0,0011 & 0,0075 & 0,0015 \\
Genre          & 0,0015 & 0,0012 & 0,0019 & 0,0015 & 0,0020 & 0,0017 & 0,0023 & 0,0021 & 0,0028 & 0,0027 \\
Actor(1)       & 0,0095 & 0,0018 & 0,0103 & 0,0024 & 0,0106 & 0,0027 & 0,0110 & 0,0033 & 0,0117 & 0,0042 \\
Writer(2)   & 0,0057 & 0,0014 & 0,0069 & 0,0022 & 0,0076 & 0,0027 & 0,0084 & 0,0032 & 0,0093 & 0,0041 \\
Director(3)    & 0,0029 & 0,0018 & 0,0036 & 0,0024 & 0,0042 & 0,0027 & 0,0050 & 0,0033 & 0,0063 & 0,0042 \\
Review         & 0,0086 & 0,0011 & 0,0095 & 0,0018 & 0,0100 & 0,0022 & 0,0105 & 0,0027 & 0,0111 & 0,0033 \\
\hline
1+2+3           & 0,0110 & 0,0016 & 0,0127 & 0,0028 & 0,0135 & 0,0036 & 0,0144 & 0,0046 & 0,0154 & 0,0058 \\
\hline
Todos           & \textbf{0,0119} & \textbf{0,0021} & \textbf{0,0133} & \textbf{0,0035} & \textbf{0,0140} & \textbf{0,0043} & \textbf{0,0148} & \textbf{0,0054} & \textbf{0,0157} & \textbf{0,0067} \\ 
\bottomrule
\end{tabular}
}
\caption{MAP values obtained by textual content based recommenders for the dataset MovieLens in Given~10 with Cosine similarity.}
\label{tab:given10_map_movielens}
\end{table}

\begin{figure}[ht] 
\centering
\includegraphics[scale=0.55]{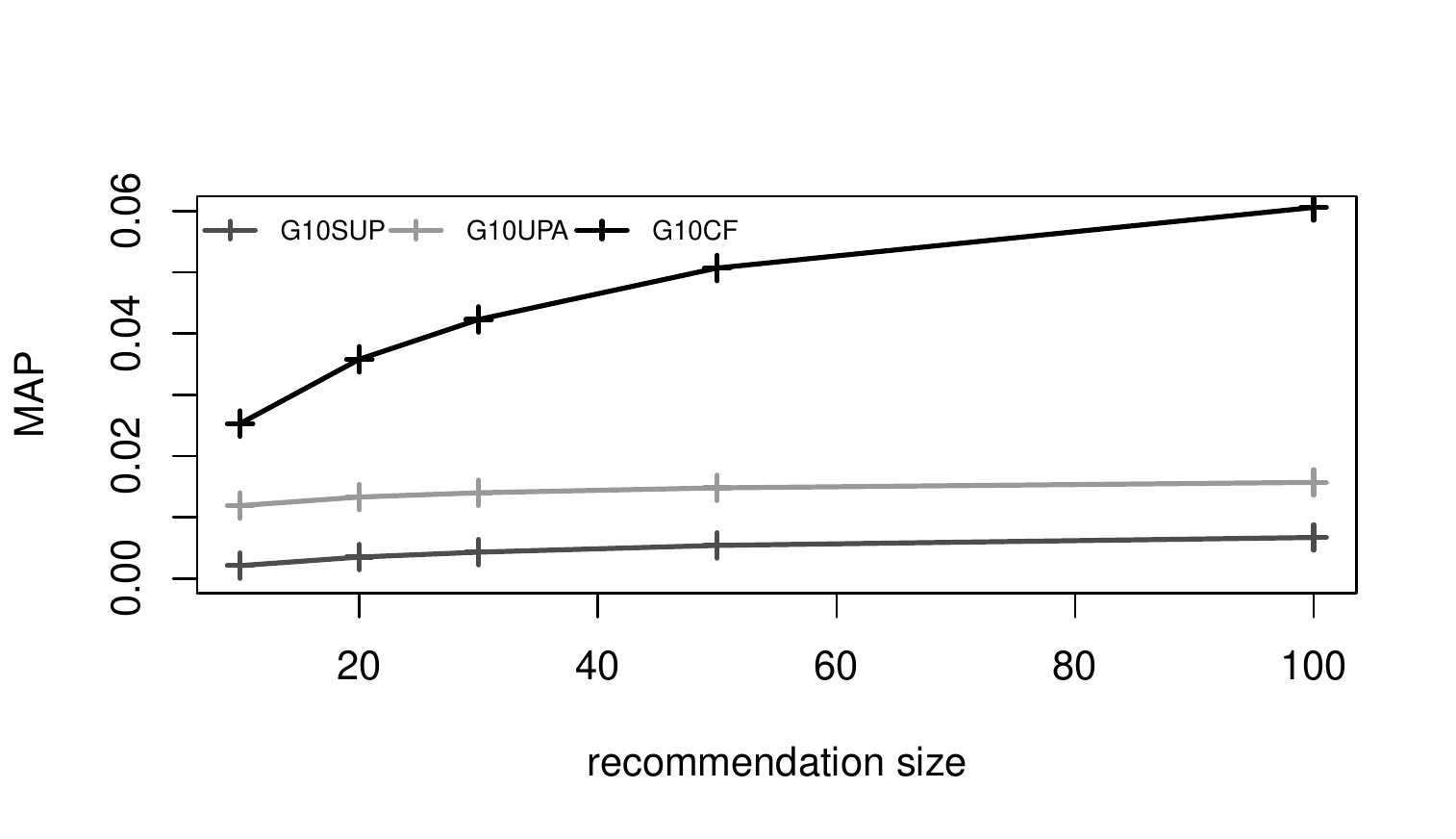}
\caption{MAP for the best results for algorithms SUP, UPA (All attributes) and CF for Movilens dataset.}
\label{img:map_ml_lines}
\end{figure}

On the other side, it is possible to observe the results for Collaborative Filtering in Movielens in Figure~\ref{img:map_ml_lines}. For any size of recommendation list, the MAP values are superior to content-based approaches. There is also a greater MAP growth rate as the size of the list increases compared to CB.

\subsubsection{Precision for CiteULike}

The  CiteULike dataset presents characteristics very different from the MovieLens. In this study, it was proposed to evaluate the same algorithms in these two datasets to observe their behavior in very different scenarios. According to Table~\ref{tab:given10_map_citeulike}, the two attributes that are most accurate for content-based recommenders using CiteULike data are \textit{Tags} and \textit{All}. However, the other fields also present interesting MAP values for this dataset. Another important aspect of the MAP values presented is that the UPA algorithm presents better MAP results comparing to the SUP algorithm, differently from that for the MovieLens dataset. However, we observed this behavior mainly in the two attributes with better MAP result, however there are variations for others.

\begin{table}[ht]
\centering
\small
\scalebox{0.65}{
\begin{tabular}{l c|c|| c|c|| c|c|| c|c|| c|c}
\toprule
& \multicolumn{10}{c}{\textbf{Given 10}} \\
\cmidrule(r){2-11}
& \multicolumn{2}{c|}{MAP@10} & \multicolumn{2}{|c|}{MAP@20} & \multicolumn{2}{|c|}{MAP@30} & \multicolumn{2}{|c}{MAP@50} & \multicolumn{2}{|c}{MAP@100}\\
\cmidrule(r){2-11}
& SUP & UPA & SUP & UPA & SUP & UPA & SUP & UPA & SUP & UPA \\
\cmidrule(r){2-11}
Title      & 0,0466 & 0,0041 & 0,0573 & 0,0212 & 0,0629 & 0,0373 & 0,0680 & 0,0519 & 0,0732 & 0,0617 \\
Abstract      & 0,0197 & 0,0135 & 0,0251 & 0,0506 & 0,0272 & 0,0701 & 0,0293 & 0,0849 & 0,0396 & 0,0936 \\
Author     & 0,0472 & 0,0161 & 0,0578 & 0,0319 & 0,0618 & 0,0438 & 0,0655 & 0,0510 & 0,0689 & 0,0544 \\
Publication  & 0,0171 & 0,0234 & 0,0244 & 0,0330 & 0,0286 & 0,0376 & 0,0328 & 0,0419 & 0,0361 & 0,0453 \\
Editor      & 0,0026 & 0,0044 & 0,0040 & 0,0053 & 0,0045 & 0,0059 & 0,0051 & 0,0067 & 0,0058 & 0,0073 \\
Source       & 0,0336 & 0,0063 & 0,0431 & 0,0155 & 0,0485 & 0,0225 & 0,0530 & 0,0292 & 0,0569 & 0,0347 \\
Tags        & \textbf{0,0796} & \textbf{0,0966} & \textbf{0,1284} & \textbf{0,1558} & \textbf{0,1509} & \textbf{0,1825} & \textbf{0,1629} & \textbf{0,1964} & \textbf{0,1694} & \textbf{0,2030} \\ 
\hline
Todos       & 0,0670 & 0,0471 & 0,1055 & 0,1093 & 0,1264 & 0,1519 & 0,1403 & 0,1714 & 0,1514 & 0,1819 \\
\bottomrule
\end{tabular}
}
\caption{MAP values obtained by textual content based recommenders for the dataset CiteULike in Given~10 with Cosine similarity.}
\label{tab:given10_map_citeulike}
\end{table}

\begin{figure}[ht] 
\centering
\includegraphics[scale=0.5]{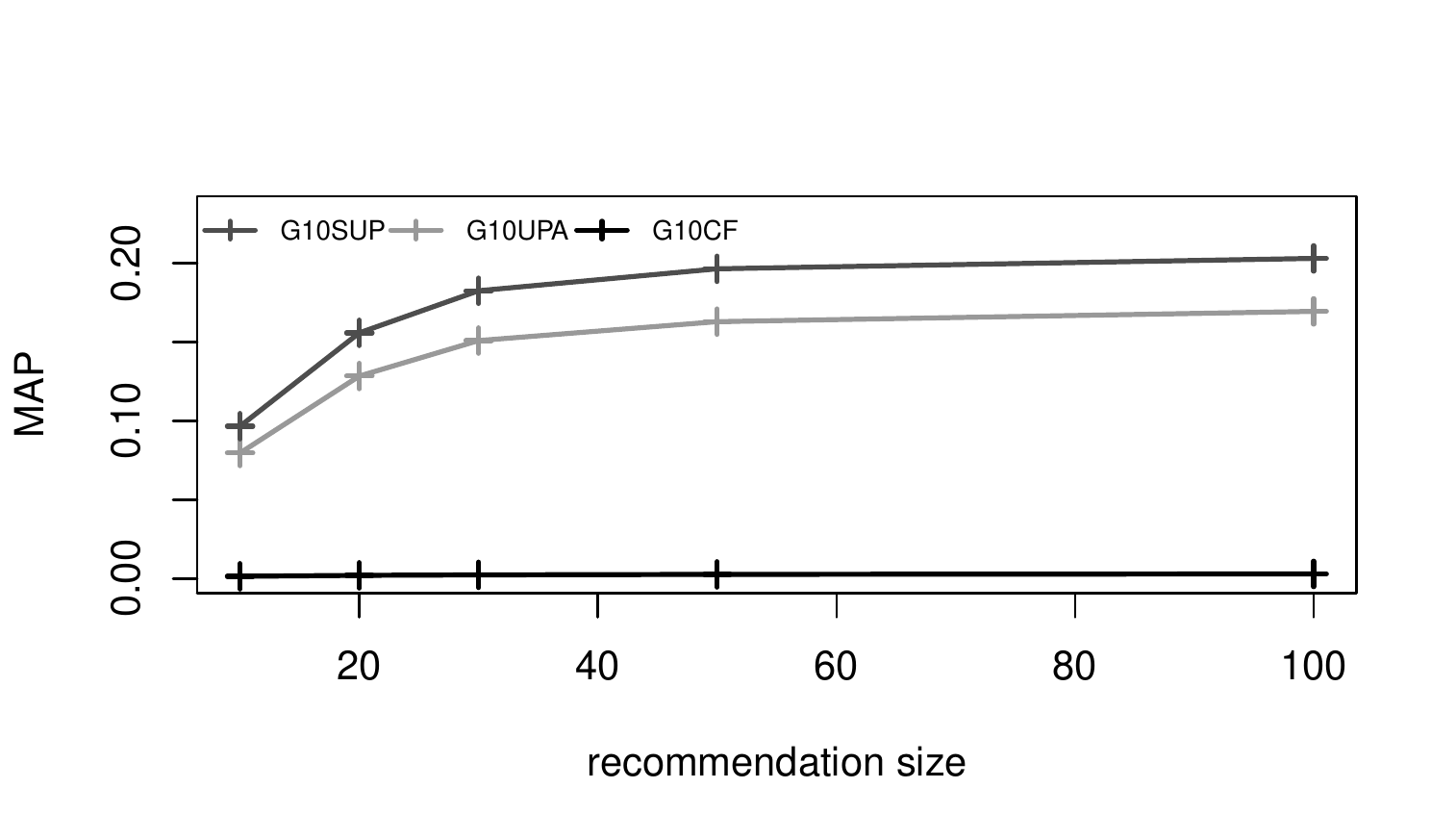}
\caption{MAP for the best results for algorithms SUP, UPA (All attributes) and CF for CiteULike dataset. }
\label{img:map_cul_lines}
\end{figure}

The dataset CiteULike has a very high sparsity in the $user \times item $ and this characteristic is determinant in the identification of the neighbors in a kNN algorithm used in Collaborative Filtering. In this scenario we expect lower MAP results for the social approach algorithm tested in this experiment as we can see in Figure~\ref{img:map_cul_lines}.

\subsection{User Coverage}

The UCOV measure indicates the ability of the system to compose a complete recommendation list for each of the users. For the CB algorithms we will present only the results obtained with the configuration that concatenates all attributes.

\subsubsection{User Coverage for MovieLens}

For all of the algorithms in any setting, the UCOV measure is $ 1.0000 $ indicating that in all configurations the recommenders are able to generate the amount of items desired for the indicated task. This indicates that no user is left without a complete list of recommended items, even for larger lists. CB found similar items and CF found similar neighbors with several distinct items.

\subsubsection{User Coverage for CiteULike}

We apply the same proposed measure of User Coverage in the CiteULike dataset as we can see in Table~\ref{tab:ucov_citeulike}. The dataset \textit{CiteULike} used in our study is a sample that aims to represent the real world as much as possible. When we use the Given~10 protocol, there are many users with only 10 or 11 items left in their test profile. In this scenario the task of ``recommending good items'' becomes very difficult, especially for social approaches when the system presents user cold-start, in which little is known about user preferences. In addition, the CiteULike system also has large item cold-start due to the large sparsity presented in the $ user \times item $ relation. This circumstance is very unfavorable for the social approach, but indifferent to the algorithms that use the textual content of the items to carry out the recommendations. The high UCOV and MAP values for this dataset gives the CB approach a better performance in this task in great contrast to the results of the CF algorithm.

\begin{table}[ht!]
\centering
\small
\begin{tabular}{l c | c | c}
\toprule
 & SUP & UPA & CF\\
\cmidrule(r){2-4}
UCOV@10  &  1,0000 &  1,0000 & 0,2474 \\ 
UCOV@20  &  1,0000 &  1,0000 & 0,2474 \\
UCOV@30  &  1,0000 &  1,0000 & 0,2398 \\
UCOV@50  &  1,0000 &  1,0000 & 0,2184 \\
UCOV@100 &  0,9864 &  1,0000 & 0,1754 \\
\bottomrule
\end{tabular}
\caption{UCOV values for CiteULike dataset.}
\label{tab:ucov_citeulike}
\end{table}

\subsection{Catalog Coverage}

Catalog Coverage aims to observe the capability of the algorithms to recover the entire catalog (all items registered in the system) among all queries performed by the system. Considering an RS that aims to help users in the problem of information overload a well-evaluated algorithm in a system test would have Precision and Coverage with the highest possible values .

\subsubsection{Catalog Coverage for MovieLens}

The catalog coverage for MovieLens, presented in Table~\ref{tab:ccov_movielens}, indicates that content-based filtering is superior to collaborative filtering in this regard. In fact, the greater precision of the collaborative approach produces limitations in coverage (or diversity) which constitutes one of the problems of social techniques as pointed out in the work of \cite{lathia2010}. Locating the balance between these two measures is a research problem not addressed in our study. However, the results for the MovieLens datset point to this question. Even if we concentrate this analysis only on the two textual content algorithms (UPA and SUP), the algorithm that presented better accuracy (SUP) has worse performance when evaluated by CCOV reaffirming this trade-off.

\begin{table}[ht!]
\centering
\small
\begin{tabular}{l c | c | c}
\toprule
& SUP & UPA & CF\\
\cmidrule(r){2-4}
UCOV@10  & 0,2024 & 0,2935 & 0,0894 \\ 
UCOV@20  & 0,3321 & 0,4420 & 0,1346 \\
UCOV@30  & 0,4285 & 0,5456 & 0,1675 \\
UCOV@50  & 0,5545 & 0,6711 & 0,2174 \\
UCOV@100 & 0,7135 & 0,8219 & 0,3018 \\
\bottomrule
\end{tabular}
\caption{CCOV values for MovieLens dataset.}
\label{tab:ccov_movielens}
\end{table}


\subsubsection{Catalog Coverage for CiteULike}

As with the values obtained in the MovieLens dataset, the CF algorithm has lower ability to cover the catalog for CiteULike. In this dataset, a purely social algorithm is hardly capable of performing the task of ``recommending good items''. As discussed from the results of precision, the large sparsity of data due to low collaboration among users makes it very difficult to locate users with similar preferences and, therefore, to identify interesting items, thus compromising not only the accuracy, but also the catalog coverage, as indicated by Table~\ref{tab:ccov_citeulike}.

Although content-based algorithms presented high values for accuracy, this did not make them inferior to CF in catalog coverage. However, coverage does not reach maximum values. The amount of items is much greater than that found in the MovieLens dataset, and with a much larger catalog, greater difficulty for high catalog coverage is expected for any algorithm. Additionally, the amount of recommendation lists generated in CiteULike is less than the number of lists generated for MovieLens due to the number of users available for testing. Even so, we could consider the CCOV results for the two algorithms SUP and UPA as expressive results  comparing the results for the two datasets for this measure.

\begin{table}[ht!]
\centering
\small
\begin{tabular}{l c | c | c}
\toprule
& SUP & UPA & CF\\
\cmidrule(r){2-4}
UCOV@10  & 0,0182 & 0,0192 & 0,0045 \\ 
UCOV@20  & 0,0357 & 0,0382 & 0,0090 \\
UCOV@30  & 0,0525 & 0,0566 & 0,0129 \\
UCOV@50  & 0,0842 & 0,0911 & 0,0195 \\
UCOV@100 & 0,1546 & 0,1666 & 0,0308 \\
\bottomrule
\end{tabular}
\caption{CCOV values for CiteULike dataset.}
\label{tab:ccov_citeulike}
\end{table}

\subsection{Similarity}

In RS's commercial solutions it is common to identify several simultaneous approaches and complement each other's shortcomings. For this reason, we believe that exploring the similarity between the results of the algorithms can provide a deeper insight into the approaches and add new questions about the differences between CF and CB.

\subsubsection{Similarity for MovieLens}

To perform the comparison between the recommendations, we generate a graph indicating, for each pair of compared algorithms, the Jaccard similarity among the recommendation lists generated by each of the two tested algorithms. Each bar chart indicates the degree of similarity to a given list size. We compared the algorithms in different sizes of lists: TOP10, TOP20, TOP30, TOP50 and Top100.

\begin{figure}[htb!] 
\centering
\includegraphics[scale=0.5]{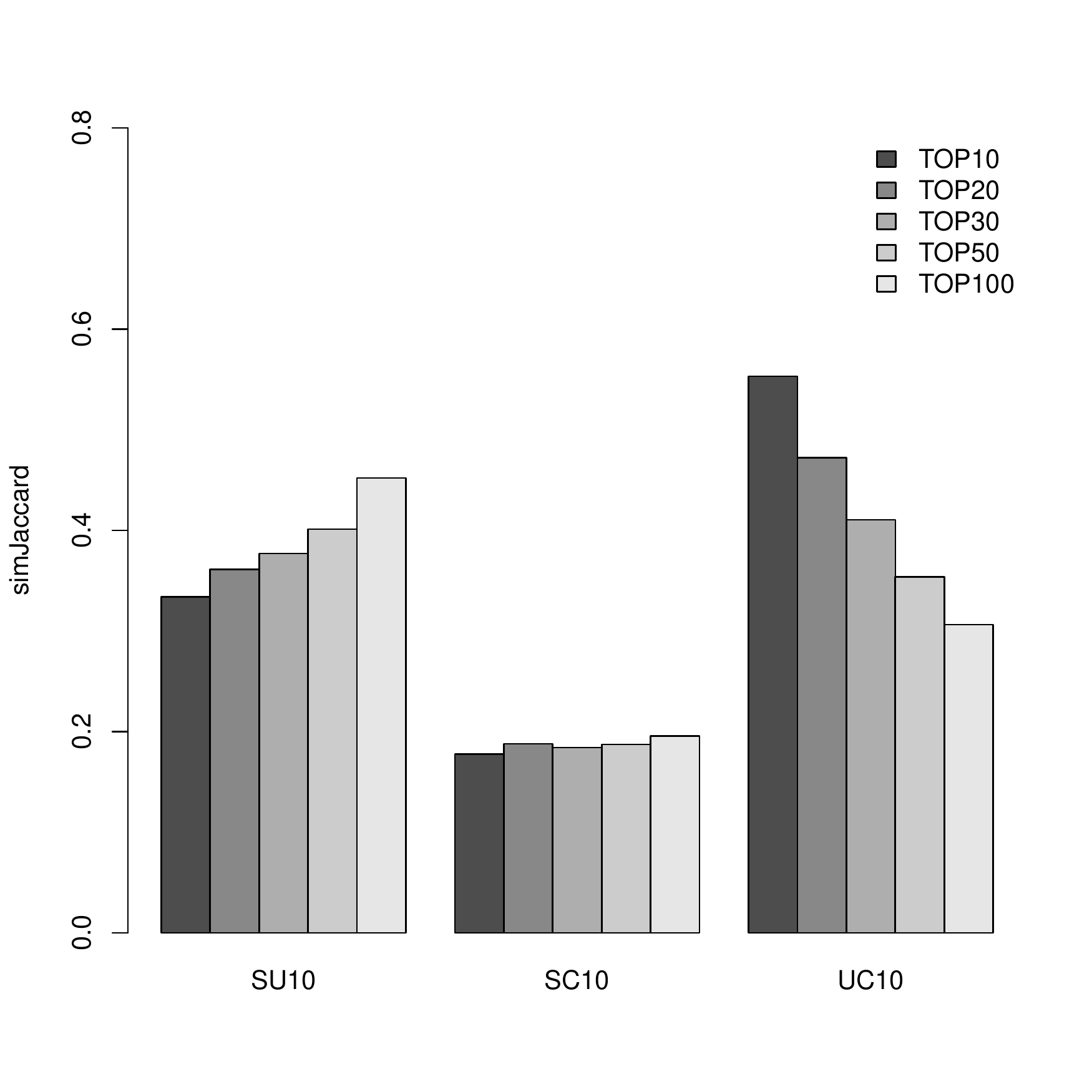}
\caption{Similarity among the recommendations generated with the MovieLens dataset.}
\label{fig:sim_ml_full}
\end{figure}

Figure~\ref{fig:sim_ml_full} highlights an expected result for the two content-based recommendation algorithms: a high degree of similarity between the recommendation lists generated by each. However, observing the superior precision results for the SUP algorithm to the UPA algorithm was expected to have a superior similarity of this algorithm in comparing the BC algorithms concerning CF. This result is not confirmed. Another interesting point is about the similarity between UPAxCF that is larger in smaller lists and decreases as the wider lists are presented to users. This behavior is opposite in the SUPxUPA similarity that grows as the recommendation lists grow in size.

\subsubsection{Similarity for CiteULike}

\begin{figure}[htb!] 
\centering
\includegraphics[scale=0.5]{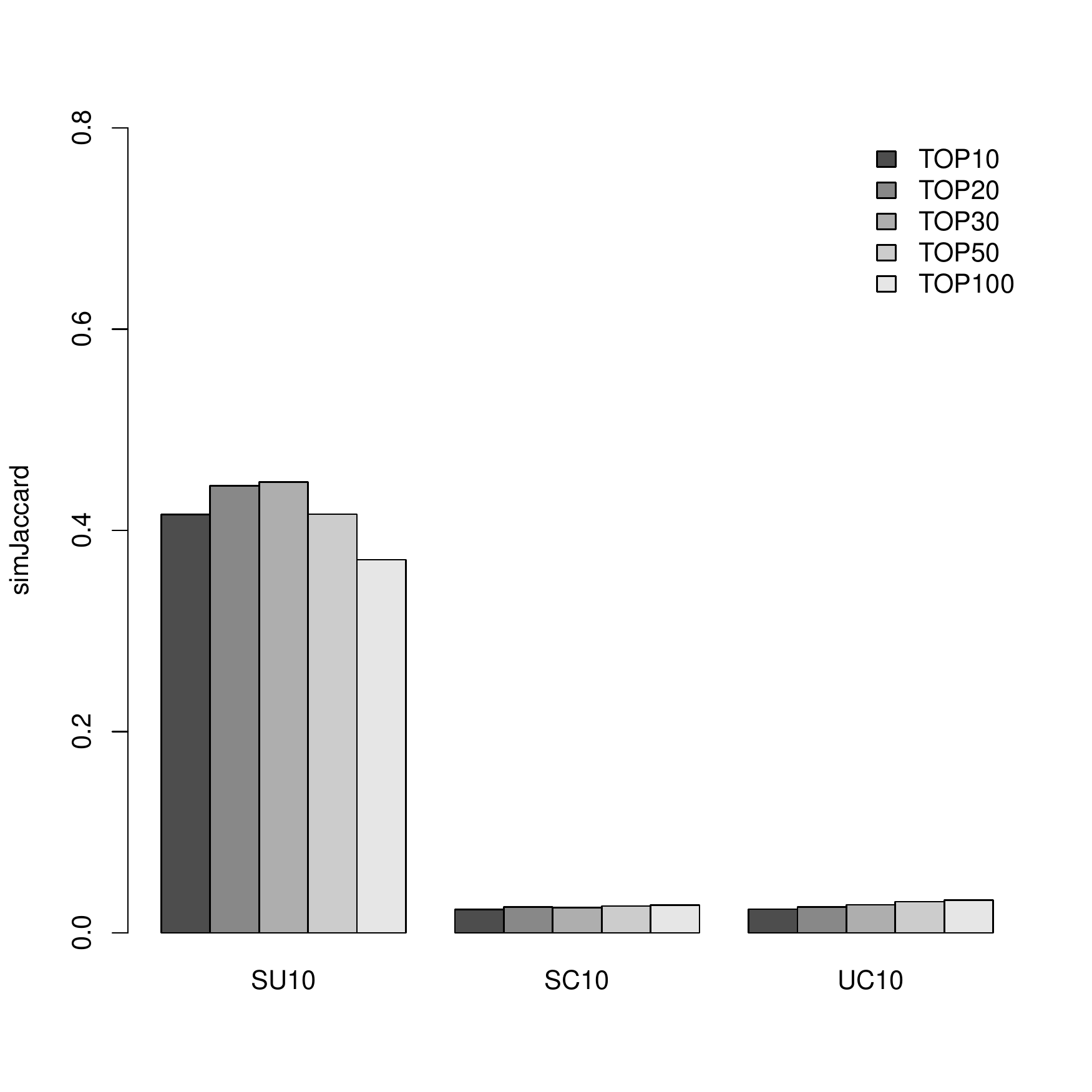}
\caption{Similarity between recommendation lists built for CiteULike dataset.}
\label{img:sim_cul_full}
\end{figure}

The same comparison performed for MovieLens was applied to the CiteULike and in Figure~\ref{img:sim_cul_full}, we present the results of similarity between the recommendation lists in this context. The comparison undertaken here suffers direct influence over the fact of CF inefficiency in ``recommending good items'' for this dataset. As already presented in the results of UCOV, this approach is not able to compose the list of recommendations entirely, and so the comparison between the lists present low similarity.

\subsection{Intersection}

In face of the results presented so far, we analyze the next aspect: do the algorithms have the same correct prediction hits?

\subsubsection{Intersection for MovieLens}

\begin{figure}[ht!]
\centering
\subfigure[SUPxCF for Given~10]{
\includegraphics[scale=0.50]{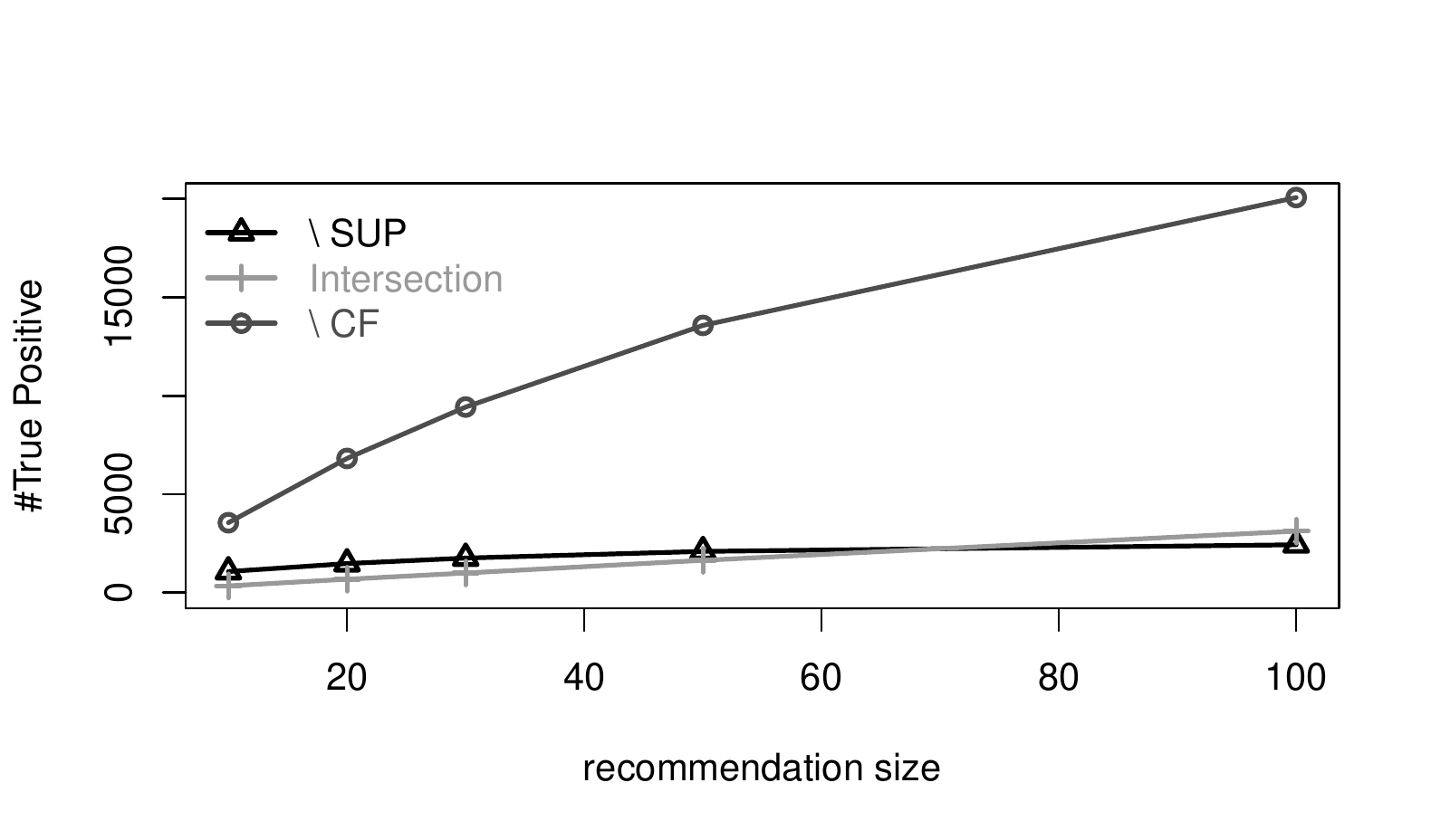}
\label{img:int_ml_g10_sc}
}
\subfigure[UPAxCF for Given~10]{
\includegraphics[scale=0.50]{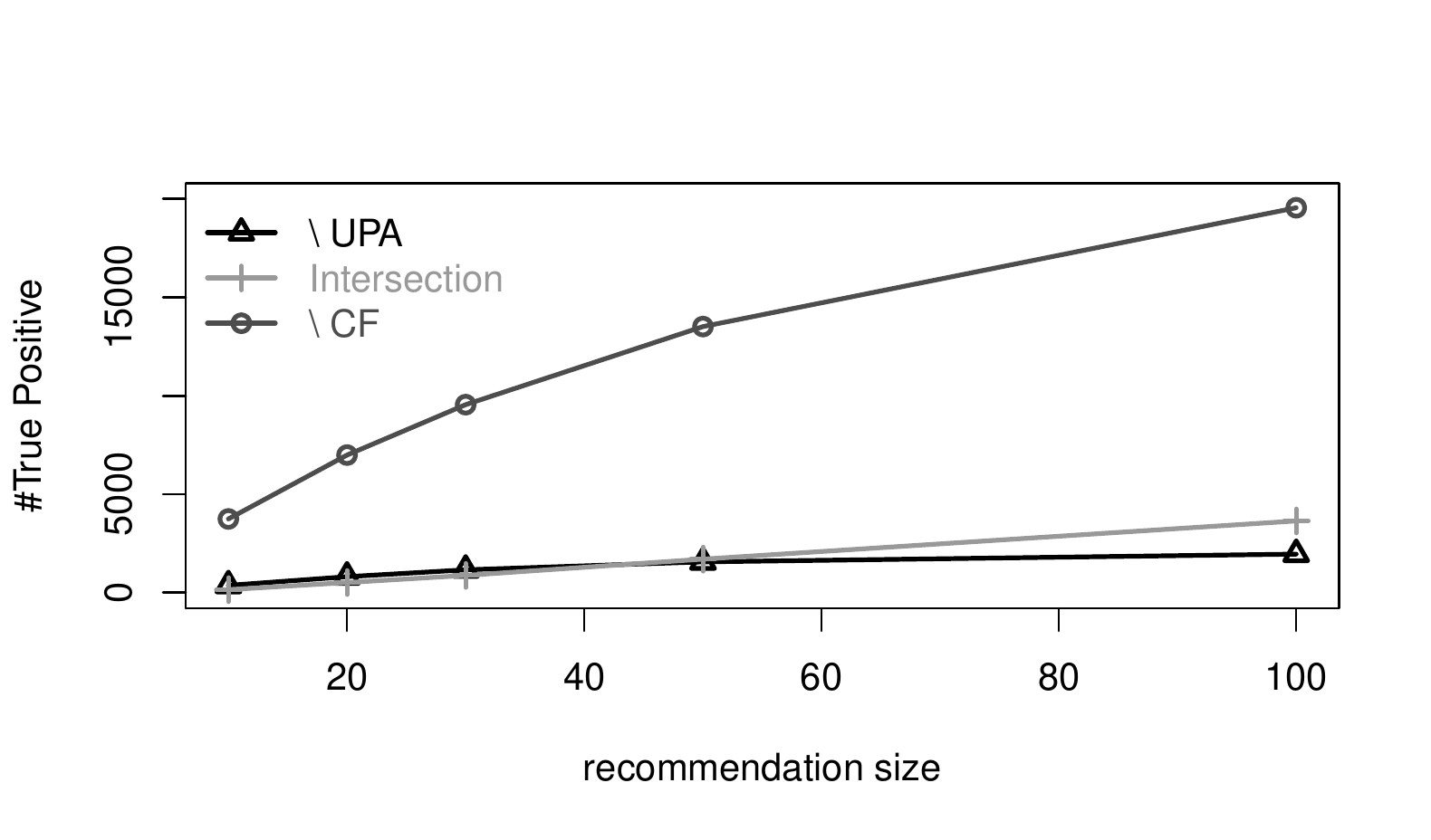}
\label{img:int_ml_g10_uc}
}
\subfigure[SUPxUPA for Given~10]{
\includegraphics[scale=0.50]{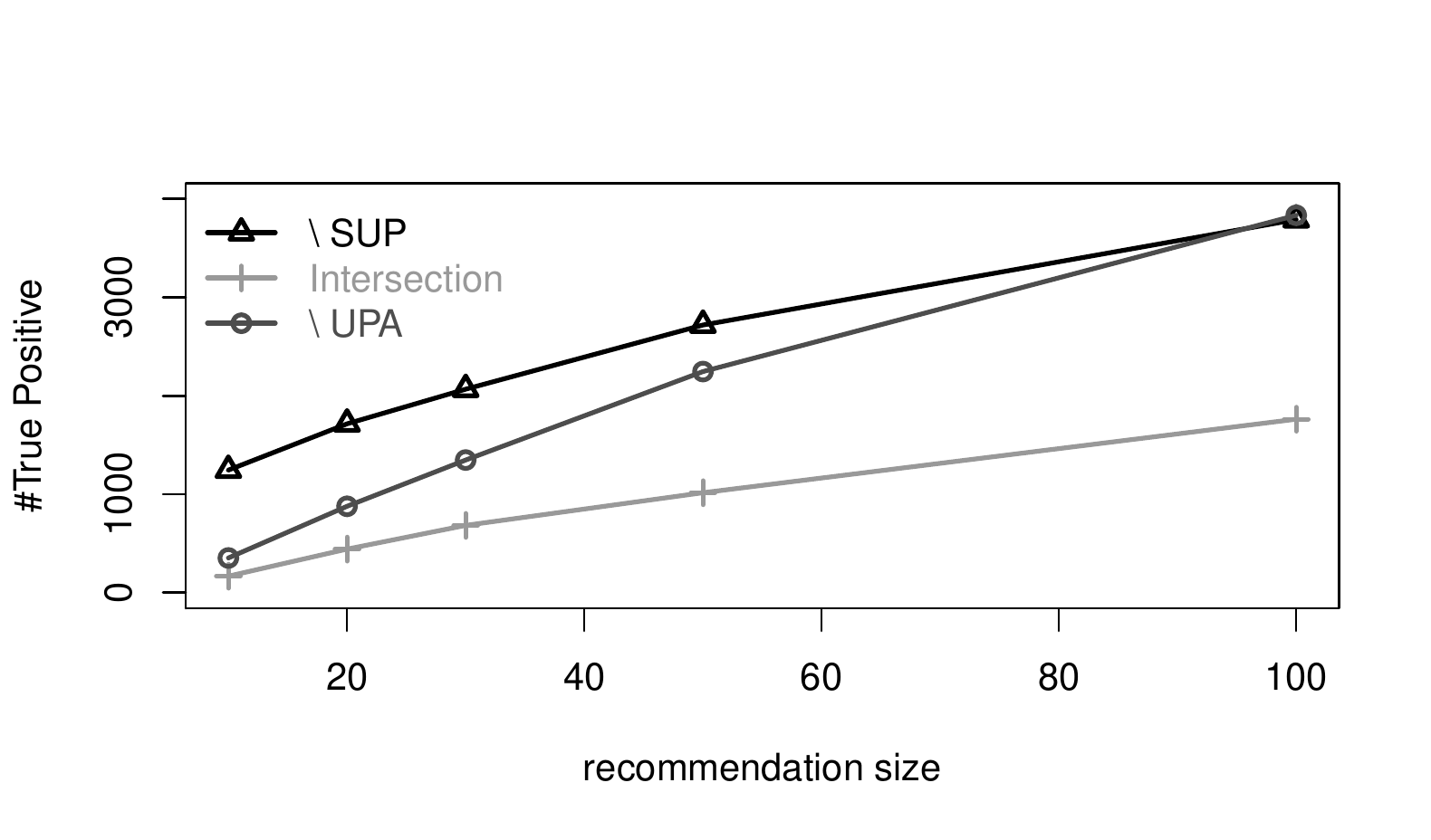}
\label{img:int_ml_g10_su}
}
\caption{Intersection among true positives for each algorithm evaluated for Given~10 for MovieLens dataset.}
\label{img:intersect_movilens}
\end{figure}

Figure~\ref{img:intersect_movilens} exhibits exclusive correct hits for each algorithm as \textit{\textbackslash~Algorithm} (e.g. \textit{\textbackslash~SUP}) and the amount of correct hits in common for both algorithms as \textit{Intersection}. All these hit intersection charts consider the recommendations generated for all test users without cross-validation. Results show that SUP and UPA algorithms produce distinct correct hits from baseline for short recommendation list. Even between SUP and UPA this occurs. Therefore, even though CF algorithm has higher precision, the predictions from SUP, in short lists are mostly different.

\subsubsection{Intersection for CiteULike}

\begin{figure}[ht!]
\centering
\subfigure[SUPxCF for Given~10]{
\includegraphics[scale=0.50]{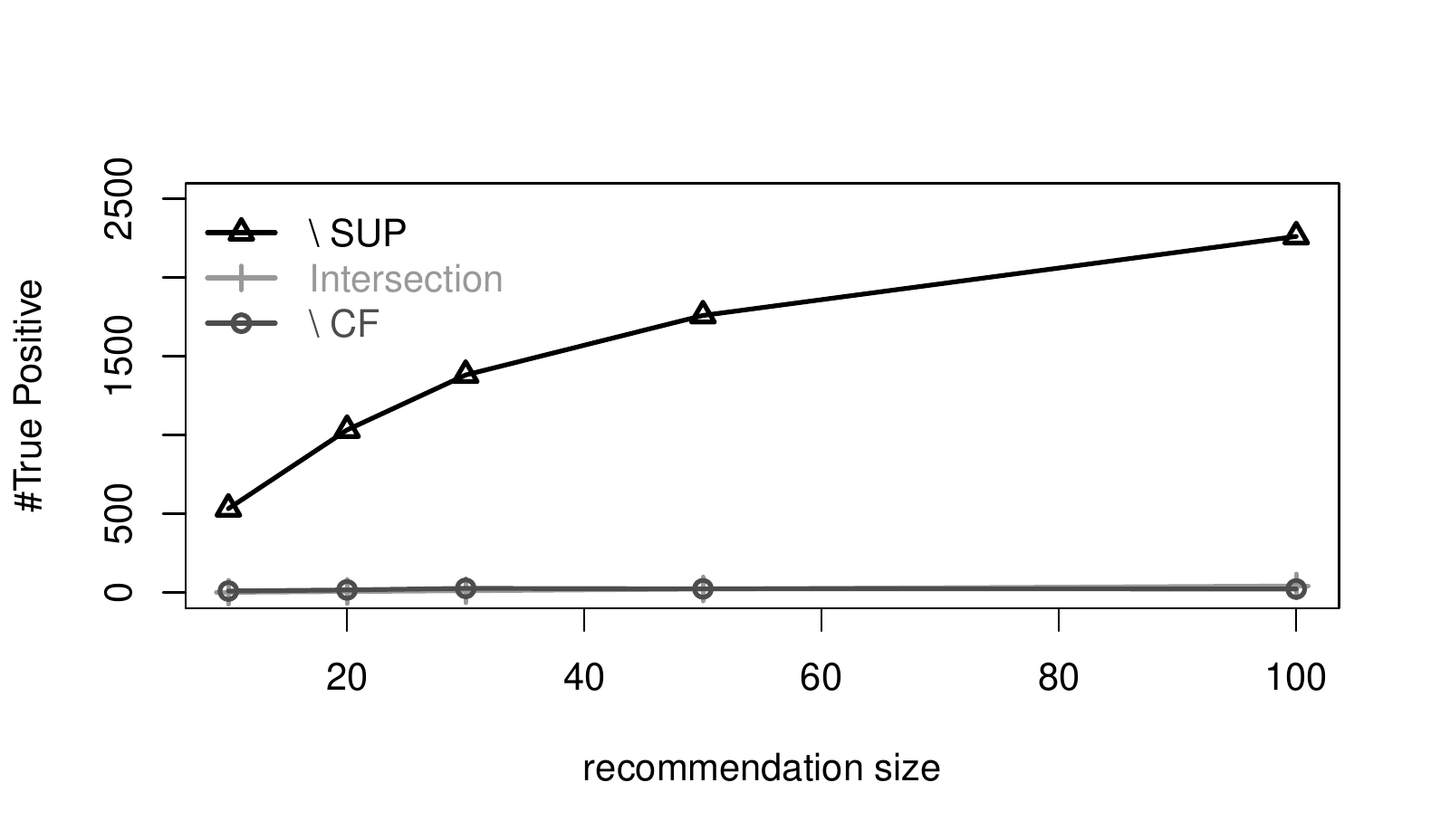}
\label{img:int_cul_g10_sc}
}
\subfigure[UPAxCF for Given~10]{
\includegraphics[scale=0.50]{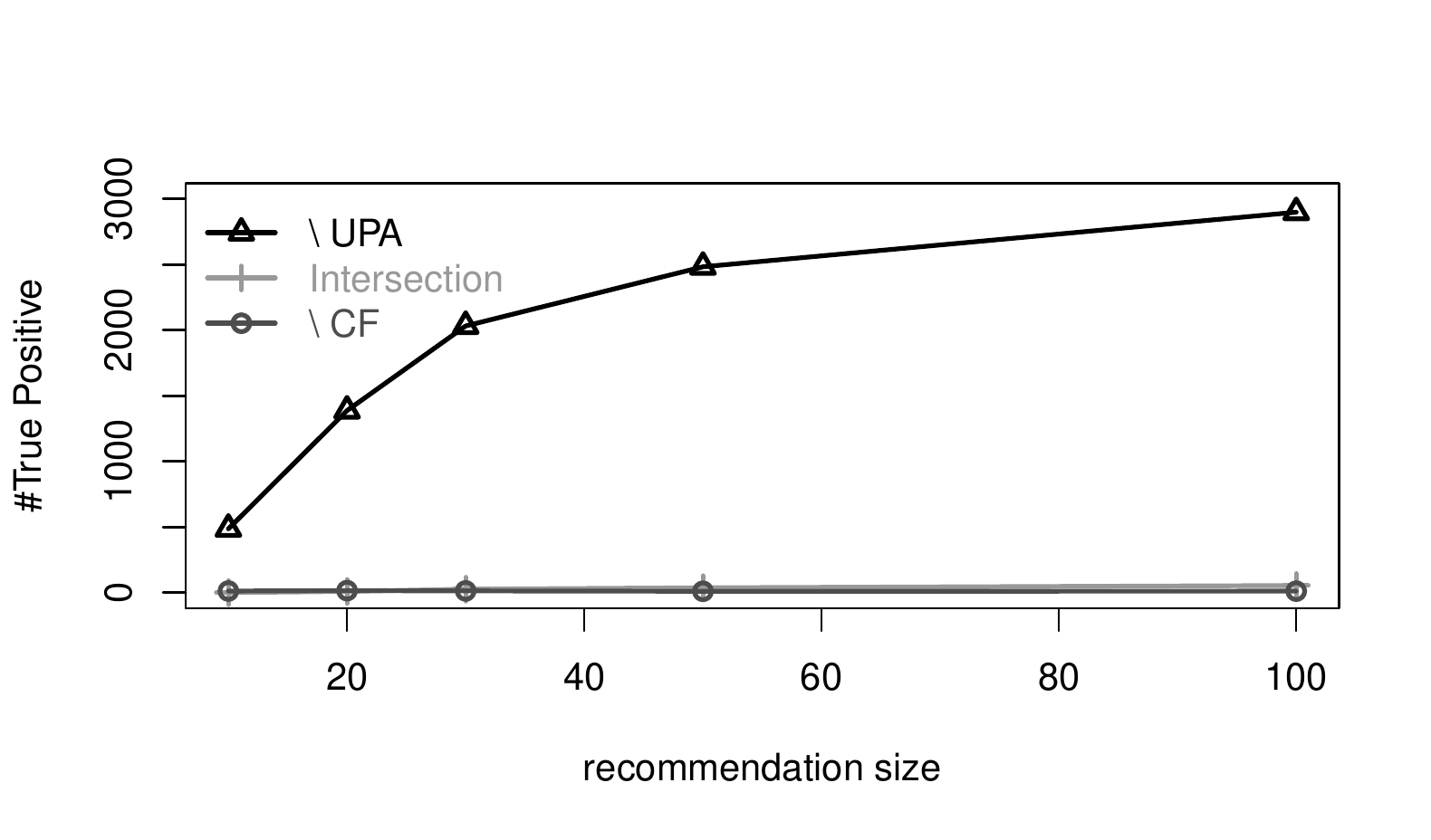}
\label{img:int_cul_g10_uc}
}
\subfigure[SUPxUPA for Given~10]{
\includegraphics[scale=0.50]{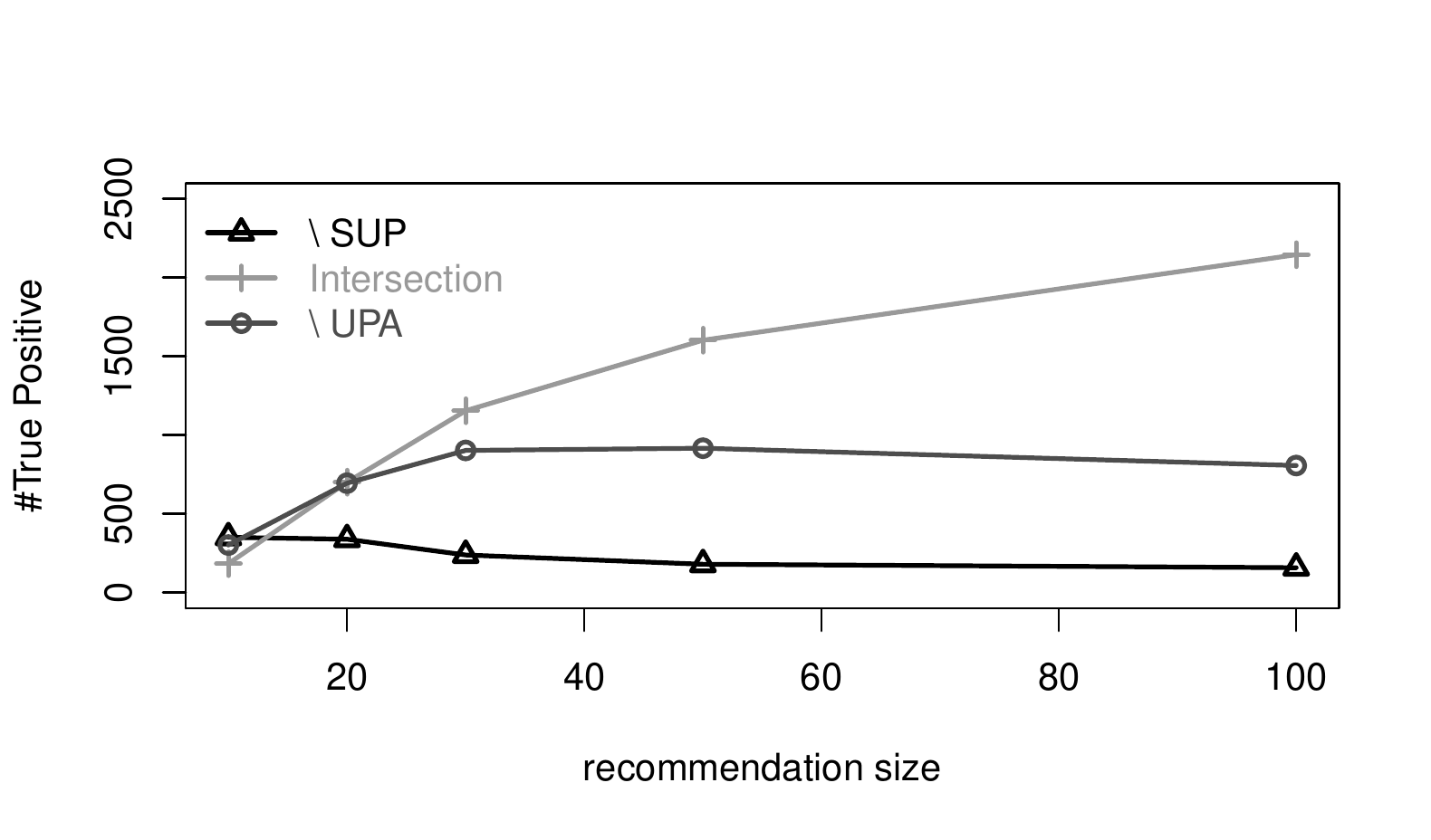}
\label{img:int_cul_g10_su}
}
\caption{Intersection among true positives for each algorithm evaluated for  Given~10 for CiteULike dataset.}
\label{img:intersect_citeulike}
\end{figure}

We present the intersection of the hits for CiteULike in Figure~\ref{img:intersect_citeulike}. We verify that the size of the recommended list of recommendations in the RS application can influence the choice of the algorithm. If the task is limited to lists of size TOP10, the SUP algorithm presents a higher rate of exclusive hits of the predictions (in addition to the high rate of hits of the intersection). However, from TOP20 the UPA algorithm presents a higher rate of exclusive hits. This behavior suggests that the SUP algorithm performs better performance in its predictions even in the first positions of the list. Even considering an unfair comparison, the social baseline presents distinct results of recommendation. So once again the predictions performed by different algorithms are different.

\section{Conclusion}
\label{sec:remarks}

The comparison results among the performed algorithms in this study help to better analyze  the differences and similarities between the two main recommendation approaches are. With the results in MovieLens, we can conclude that the SUP algorithm has an exciting feature: its predictions are mostly different to those performed by CF in smaller recommendation lists. When the size of the recommended lists increases, then the results of CB are overlapping by CF. Besides, our results clarify the proposition raised in the study of \cite{burke2002} that discusses the complementary aspects of the recommendation approaches. Evaluating the task of ``recommending good items'', in addition to checking the catalog coverage with superior results for BC, we also find that the predictions are different from that of a typical social recommender.

When we run two CB algorithms, we are expecting that the higher the volume of data available the better the performance of these algorithms would be. This behavior exists in the MAP results for the \textit{All} attribute that concatenates all the attributes retrieved from the IMDb. However, the results presented by \textit{Actor} and the concatenation of \textit{Actor + Writers + Directors} are expressive results for this approach. This result suggests that the contents of the items have the different relevance. The low results for the \textit{Gender} attribute is an example. A user should not like a particular movie because it has the same gender as another well-evaluated one previously. Besides, there are many movies with the same genre making it difficult to filter because it is not possible to distinguish the movies. Opposite to this situation, the good results for the \textit{Actors} attribute suggest that many users enjoy several movies starring the same actors. These MAP results presented by the SUP algorithm, although smaller to those presented by FC, denotes that it is possible to perform the task of ``recommending good items'' through item content. However, the choice of what content to use to carry out the recommendations is an important decision in the recommendation's design.

In contrast to what occurred in the MAP results in the MovieLens dataset, the values presented for CiteULike demonstrate the inefficiency of social approaches in such context: a large number of items and low collaboration among users. When analyzing  Figure~\ref{img:map_cul_lines} the problem becomes clear. In this case, the social techniques are difficult to apply, because to locate users with similar interests (with the same publications) would imply in isolating users in co-citations groups that restricting the possibilities of recovery of new items. 

\subsection{Future work}

We believe that a great contribution from our work is to present through a detailed observation, the complementary feature between algorithms of the two main recommendation approaches. It is possible to realize that there is relevant and exclusive content between the recommendation lists of the two approaches. Therefore, a next step would be to expand this study to how to evaluate better the combinations of the results of these algorithms by exploring this complimentary feature of the algorithms in hybrid systems \cite{glauber2013mixed, aslanian2016hybrid, liu2018novel}. 

Our attempt to generalize the results we tested the recommendation algorithms proposed in two different datasets: MovieLens+IMDb and CiteULike. However, there are several other types of items that overwhelm users in their Web experiences. Contents such as news, book, and music are some within the diversity existing on the Web. New offline tests for a more significant amount of content is necessary and should be explored next. Another important aspect is to expand the experiments for other recommendation algorithms and different approaches.

In recommendation algorithm projects, it is common to define two test steps: offline and online tests. Our research focused on the first step in which there is no need for the presence of users. Building a platform for the proposed recommendation algorithms to be asked to present their recommendations to real users is another step in this research. Some studies such as those of \cite{konstan2012} and mainly in the research of \cite{mcnee2006being} are presented the perspective that not always the highest accuracy in the forecasts corresponds to the greater satisfaction of the users. In this case, considering the RS as a tool to support a better user experience on the Web, minimizing the problem of information overload, an assessment able to measure their degree of satisfaction is of great importance in a project of RS.

\bibliographystyle{ACM-Reference-Format}
\bibliography{sigproc} 

\end{document}